\documentclass[11pt]{article}
\usepackage{epsfig}

\begin{document}

\title{{\bf{\Large Voros product and the Pauli 
principle at low energies}}}
\author{
{\bf {\normalsize Sunandan Gangopadhyay}$^{a,b,}
$\thanks{sunandan.gangopadhyay@gmail.com, sunandan@bose.res.in}},\,
{\bf {\normalsize Anirban Saha}$^{a,c}
$\thanks{anirban@bose.res.in}}\\
{\bf {\normalsize Frederik. G. Scholtz}
$^{d,}$\thanks{fgs@sun.ac.za}}\\
$^{a}$ {\normalsize Department of Physics, West Bengal State University, Barasat, India}\\
$^{b}${\normalsize Visiting Associate in S.N. Bose National Centre for Basic Sciences,}\\
{\normalsize JD Block, Sector III, Salt Lake, Kolkata-700098, India}\\[0.3cm]
$^{c}${\normalsize Visiting Associate in Inter University Centre for Astronomy and Astrophysics,}\\
{\normalsize Ganeshkhind, Pune, India}\\[0.3cm]
$^{d}$ {\normalsize National Institute of Theoretical Physics, 
Stellenbosch,}\\{\normalsize Stellenbosch 7600, South Africa}\\[0.3cm]
}
\date{}

\maketitle


\begin{abstract}

\noindent Using the Voros star product, we investigate the status of the 
two particle correlation function to study the possible 
extent to which the previously proposed violation of the Pauli 
principle may impact at low energies. The results show interesting
features which are not present in the computations made using the Moyal
star product.


\\[0.3cm]

\end{abstract}


\noindent It is likely that at short distances spacetime has to be
described by different geometrical structures and that the very concept
of point and localizability may no longer be adequate. This is one of the
main motivations for the introduction of noncommutative (NC) geometry \cite{connes}-\cite{bondia}
in physics. Attempts have been made to formulate quantum mechanics
\cite{mezincescu}-\cite{fgs} and quantum field theories \cite{seiberg}-\cite{doug} on NC spacetime.
However, for the latter case, the issue of the lack of Lorentz invariance
symmetry has remained a challenge since field theories defined on a NC spacetime
with the commutation relation of the coordinate operators
\begin{eqnarray}
[\hat{x}_{\mu}, \hat{x}_{\nu}]&=& i\theta_{\mu\nu}
\label{eq1}
\end{eqnarray}
where $\theta_{\mu\nu}$ is a constant antisymmetric matrix, are obviously not
Lorentz invariant.

The twist approach to NC quantum field theory has been developed to circumvent
this problem \cite{chaichian}-\cite{bal1}. It is triggered by the realization that it is possible
to twist the coproduct of the universal envelope $U(\mathcal{P})$ of the Poincar\'e
algebra such that it is compatible with the star product.

An interesting result that follows from the twisted implementation of the Poincar\'e group is that the Bose and Fermi commutation relations get deformed \cite{balachandran}. This is controversial though as a different point of view was presented in \cite{wess} and the issue is not yet resolved. A further controversy involves the use of the Moyal or Voros twist, related to the use of the Moyal or Voros star product \cite{bal2,tanasa,lizzi1}. In this article we adopt the point of view of \cite{balachandran} and consider the effect of implementing the twist through the Voros or Moyal twist within a simple nonrelativistic setting.  

A very striking consequence of these deformations is the violation of the Pauli exclusion principle.  In \cite{sgfgs}, it has been shown that the two particle correlation function for a free fermion gas in $2+1$ dimensions (with exclusively spatial noncommutativity, i.e. $\theta_{0i}=0$, $i=1,2$)  does not vanish in the limit $r\rightarrow 0$ which indicates that there is a finite probability that fermions may come very close to each other. The computation has been done by using twisted commutation relations among the creation and annihilation operators of the Schr\"{o}dinger field which in turn has been obtained by taking the non-relativistic limit of the analogous relations for the Klein-Gordon field. A key ingredient in the entire analysis has been the Moyal twist element which reads
\begin{eqnarray}
\mathcal{F}^{(\theta)}_{M}&=&e^{-\frac{i}{2}\theta^{ij}\partial_{i}\otimes \partial_{j}}
\label{moyal}
\end{eqnarray} 
where (in two spatial dimensions) $\theta^{ij}=\theta\epsilon^{ij}$ with $\epsilon^{ij}$ being the antisymmetric tensor of order two.
 
\noindent Recently, a comparison of NC field theories built on two 
different star products (twist elements), namely the Moyal and Voros, have been made \cite{lizzi1}. 
It has been found that although the Green's functions
are different for the two theories, the S-matrix is the same in both cases and is
different from the commutative case. This motivates us to investigate the status
of the two particle correlation function using deformed commutation relations
obtained by incorporating the action of the Voros twist element \cite{lizzi1}\footnote{The Voros product has also
played a key role in the obtention of a $\theta$-deformed mass density of a spherically symmetric gravitational source \cite{sgasfgs}.} 
\begin{eqnarray}
\mathcal{F}^{(\theta)}_{V}&=&e^{-\theta\partial_{+}\otimes \partial_{-}}\\
\partial_{\pm}&=&\frac{1}{\sqrt{2}}\left(\frac{\partial}{\partial x^{1}}\mp i\frac{\partial}{\partial x^{2}}\right)\nonumber
\label{voros}
\end{eqnarray}   
on the usual pointwise product between two fields.

To begin the analysis, we first write down the mode expansion 
of a free nonrelativistic quantum field $\psi$ of mass $m$ (in units with $\hbar=1$)
in the NC plane as
\begin{eqnarray}
\psi(\vec{r} , t) = \int d^{2}\vec{k}~a(\vec{k})e_{k}(\vec{r})
\label{mode_exp}
\end{eqnarray}
where $e_{k}(\vec{r})=e^{-i\frac{|\vec{k}|^{2}t}{2m}}e^{i\vec{k}.\vec{r}}$
and $a(\vec{k})$ satisfy the usual (anti)commutation relation
\begin{eqnarray}
a(\vec{k}) a^{\dagger}(\vec{k}')-\eta a^{\dagger}(\vec{k}')a(\vec{k})=(2\pi)^{2}\delta^{2}(\vec{k}-\vec{k}')
\label{commutation}
\end{eqnarray}
where $\eta$ is $+1$ for bosons and $-1$ for fermions.
The deformation algebra involving $a(\vec{k})$ has already been
derived in \cite{lizzi} for the relativistic case and can be readily 
shown to hold in the nonrelativistic case as well :
\begin{eqnarray}
a(\vec{k})\star_{M}a(\vec{k}')&=&e^{-(1/2)\theta^{ij}k_{i}k'_{j}}a(\vec{k})a(\vec{k}')\nonumber\\
a(\vec{k})\star_{M}a^{\dagger}(\vec{k}')&=&e^{(1/2)\theta^{ij}k_{i}k'_{j}}a(\vec{k})a^{\dagger}(\vec{k}')\nonumber\\
a^{\dagger}(\vec{k})\star_{M}a(\vec{k}')&=&e^{-(1/2)\theta^{ij}k_{i}k'_{j}}a^{\dagger}(\vec{k})a(\vec{k}')
\label{deform_moyal}
\end{eqnarray}
\begin{eqnarray}
a(\vec{k})\star_{V}a(\vec{k}')&=&e^{-\theta k_{-}k'_{+}}a(\vec{k})a(\vec{k}')\nonumber\\
a(\vec{k})\star_{V}a^{\dagger}(\vec{k}')&=&e^{\theta k_{-}k'_{+}}a(\vec{k})a^{\dagger}(\vec{k}')\nonumber\\
a^{\dagger}(\vec{k})\star_{V}a(\vec{k}')&=&e^{-\theta k_{-}k'_{+}}a^{\dagger}(\vec{k})a(\vec{k}')
\label{deform_voros}
\end{eqnarray}
where $k_{\pm}=\frac{1}{\sqrt{2}}(k_{1}\pm ik_{2})$.


\noindent We now compute the two particle correlation 
function for a free gas in 2+1 dimensions using the 
canonical ensemble, i.e., we are interested in 
the matrix elements $\frac{1}{Z}\langle \vec{r}_1, \vec{r}_2|e^{-\beta H}|
\vec{r}_1, \vec{r}_2\rangle$, ($\beta=1/(k_{B}T)$) where $Z$ is the canonical partition function
and $H$ is the non-relativistic Hamiltonian.  
The physical meaning of this function is quite simple; 
it tells us what the probability is to find particle two 
at position $\vec{r}_2$, given that particle one is at $\vec{r}_1$, i.e., 
it measures two particle correlations \cite{pathria}. The relevant two 
particle state is given by
\begin{eqnarray}
|\vec{r}_{1}, \vec{r}_{2}\rangle&=&
\hat\psi^{\dag}(\vec{r}_{1})\star_{V}\hat\psi^{\dag}(\vec{r}_{2})|0\rangle\nonumber\\
&=&\int \frac{d^{2}\vec{q}_{1}}{(2\pi)^2}\frac{d^{2}\vec{q}_{2}}{(2\pi)^2}
e^{*}_{q_{1}}(\vec{r}_{1})e^{*}_{q_{2}}(\vec{r}_{2})a^{\dag}(\vec{q}_{1})\star_{V}a^{\dag}(\vec{q}_{2})
|0\rangle\,.
\label{rh1}
\end{eqnarray}
The two particle correlation function can therefore be written as
\begin{eqnarray}
C(\vec{r}_{1}, \vec{r}_{2})=\frac{1}{Z}\langle \vec{r}_{1}, \vec{r}_{2}|e^{-\beta H}|\vec{r}_{1}, \vec{r}_{2}\rangle
=\frac{1}{Z}\int d^{2}\vec{k}_{1}d^{2}\vec{k}_{2}e^{-\frac{\beta}{2m}(\vec{k}^{2}_{1}+\vec{k}^{2}_{2})}
|\langle \vec{r}_{1}, \vec{r}_{2}|\vec{k}_{1}, \vec{k}_{2}\rangle|^{2}
\label{rh2}
\end{eqnarray}
where we have introduced a complete set of momentum 
eigenstates $|\vec{k}_{1}, \vec{k}_{2}\rangle$ and $Z$ is the partition function of the system.

\noindent Using eq(s)(\ref{rh1}, \ref{deform_voros}, \ref{commutation}) and noting that
\begin{eqnarray}
|\vec{k}_{1}, \vec{k}_{2}\rangle=a^{\dag}(\vec{k}_1)\star_{V}a^{\dag}(\vec{k}_2)|0\rangle
\label{rh3}
\end{eqnarray}
we get
\begin{eqnarray}
|\langle \vec{r}_{1}, \vec{r}_{2}|\vec{k}_{1}, \vec{k}_{2}\rangle|^2&=&
e^{-2\theta(k_{1x}k_{2x}+k_{1y}k_{2y})}\{2+\eta e^{i(\vec{k_1}-\vec{k_2}).\vec{r}_{12}}
e^{i\theta k_{1}\wedge k_{2}} + c.c\}
\label{overlap}
\end{eqnarray}
where $\vec{r}_{12}=\vec{r}_1 -\vec{r}_2$, 
$\theta k_{1}\wedge k_{2}=\theta\epsilon^{ij}k_{1i}k_{2j}=\theta(k_{1x}k_{2y}-k_{1y}k_{2x})$ 
and $c.c$ implies complex conjugate of the second term in the above expression.
Substituting this in eq(\ref{rh2}), we obtain
\begin{eqnarray}
C(\vec{r}_{1}, \vec{r}_{2})&=&\frac{f(\theta)}{Z}
\left(1+\eta \frac{(1-\frac{4m^{2}\theta^{2}}{\beta^{2}})}{(1-\frac{3m^{2}\theta^{2}}{\beta^{2}})}
e^{- \frac{m(1+\frac{2m\theta}{\beta})}{\beta(1-\frac{3m^{2}\theta^{2}}{\beta^{2}})}r^2}\right)
\label{rh4}
\end{eqnarray}
where $r=|\vec{r}_{1}-\vec{r}_{2}|$ and $f(\theta)=\frac{2(2m\pi/\beta)^2}{1-4m^2 \theta^2/\beta^2}$.
The partition function of the system can now be readily computed and reads
\begin{eqnarray}
Z&=&\int d^{2}\vec{r}_{1}d^{2}\vec{r}_{2}\langle \vec{r}_{1}, \vec{r}_{2}|e^{-\beta H}|\vec{r}_{1}, \vec{r}_{2}\rangle\nonumber\\
&=&f(\theta)A^{2}\left\{1+\eta\frac{\pi\beta}{mA}\left(1-\frac{2m\theta}{\beta}
\right)\right\}\approx f(\theta)A^{2}
\label{partition_function}
\end{eqnarray}
in the limit of the area of the system $A\rightarrow\infty$.
Substituting the above result in eq(\ref{rh4}), we finally obtain
\begin{eqnarray}
C(\vec{r}_1, \vec{r}_2)&=&\frac{1}{A^2}
\left(1+\eta \frac{(1-\frac{16\pi^{2}\theta^{2}}{\lambda^{4}})}{(1-\frac{12\pi^{2}\theta^{2}}{\lambda^{4}})}
e^{- \frac{2\pi(1+\frac{4\pi\theta}{\lambda^2})}{\lambda^2(1-\frac{12\pi^{2}\theta^{2}}{\lambda^{4}})}r^2}\right)
\label{probability}
\end{eqnarray}
where $\lambda$ is the mean thermal wavelength given by
\begin{eqnarray}
\lambda&=&\left(\frac{2\pi\beta}{m}\right)^{1/2}\quad;\quad
\beta=\frac{1}{k_{B}T}~.
\label{rh5}
\end{eqnarray}
We quote the corresponding result in the Moyal case \cite{sgfgs} in order to compare with the Voros result obtained above (\ref{probability}):
\begin{eqnarray}
C(\vec{r}_1, \vec{r}_2)&=&\frac{1}{A^2}
\left(1+\eta \frac{1}{1+\frac{\theta^{2}}{\lambda^4}}
e^{- \frac{2\pi }{\lambda^{2}(1+\frac{\theta^2}{\lambda^4})}r^2}\right)~.
\label{probability_moyal}
\end{eqnarray}



\noindent Eq(\ref{probability}) shows many interesting features. Note that there exists two thermal wavelengths $\lambda_{1}=2\sqrt{\pi\theta}$ and $\lambda_{2}=\sqrt{2\sqrt{3}\pi\theta}$ where the second term in the above expression vanishes. These wavelengths correspond to two temperatures $T_{1}=1/(2k_{B}m\theta)$ and $T_{2}=1/(\sqrt{3}k_{B}m\theta)$ at which the correlation function is completely independent of whether the particles are bosons or fermions. For temperatures much less than $T_1$ the correlations exhibit mild deviations away from that of commutative bosons or fermions.  However, for temperatures $T$ lying between $T_1$ and $T_2$ (i.e. $T_1 < T < T_2$), the bosons start behaving as fermions and vice versa. These temperatures are of course extremely high.  Indeed, assuming $\theta$ to be of the order of the Planck length squared and restoring $\hbar$ one finds them to be of the order of $10^{48}$K.  For temperatures $T > T_2$, the exponential in eq(\ref{probability}) becomes positive and correlation grow exponentially with distance.  This signals the onset of instability, which can best be seen if one considers the exchange potential defined by $V(r)=-k_{B}T\log C(r_{1}, r_{2})$. This becomes unstable or complex above this temperature. This suggests the existence of a high energy cut off $E=k_{B}T_{2}$.  These features are completely absent when the computations are made using the Moyal star product \cite{sgfgs} and are summarized in Figures 1a, 1b, 2a and 2b, which shows the exchange potential for the Moyal and Voros twist in the case of bosons and fermions, respectively, for values of the dimensionless variable $\alpha=\frac{\theta}{\lambda^2}$ below $T_1$ ($\alpha=0.0796$) and between $T_1$ and $T_2$ ($\alpha=0.919$).  Here $r$ is measured in units of $\lambda$.  For reference the commutative results ($\alpha=0$) are also shown in which case there is of course no distinction between Moyal and Voros.  The transgression of statistics in the temperature range between $T_1$ and $T_2$ can clearly be seen.  Of course, if these temperatures are really as high as advertised here, the nonrelativistic limit would be invalid and the conclusions above will probably be modified in a proper relativistic description.  However, the crucial point here is not so much the absolute values of $\theta$ and $\lambda$, but rather the existence of two length scales ($\sqrt{\theta}$ and $\lambda$), which is not the case in commutative theories, and the (not unexpected) non-trivial behaviour that occurs when these two length scales become comparable. This may be quantitatively different in a relativistic treatment, but generically one would again expect non-trivial behaviour when the two length scales are comparable. In effective noncommutative systems this may happen at much lower temperatures where the nonrelativistic approximation is valid.
     
\setlength{\unitlength}{1mm}
\begin{figure}
\begin{picture}(53,53)
\put(0, 0){\epsfig{file=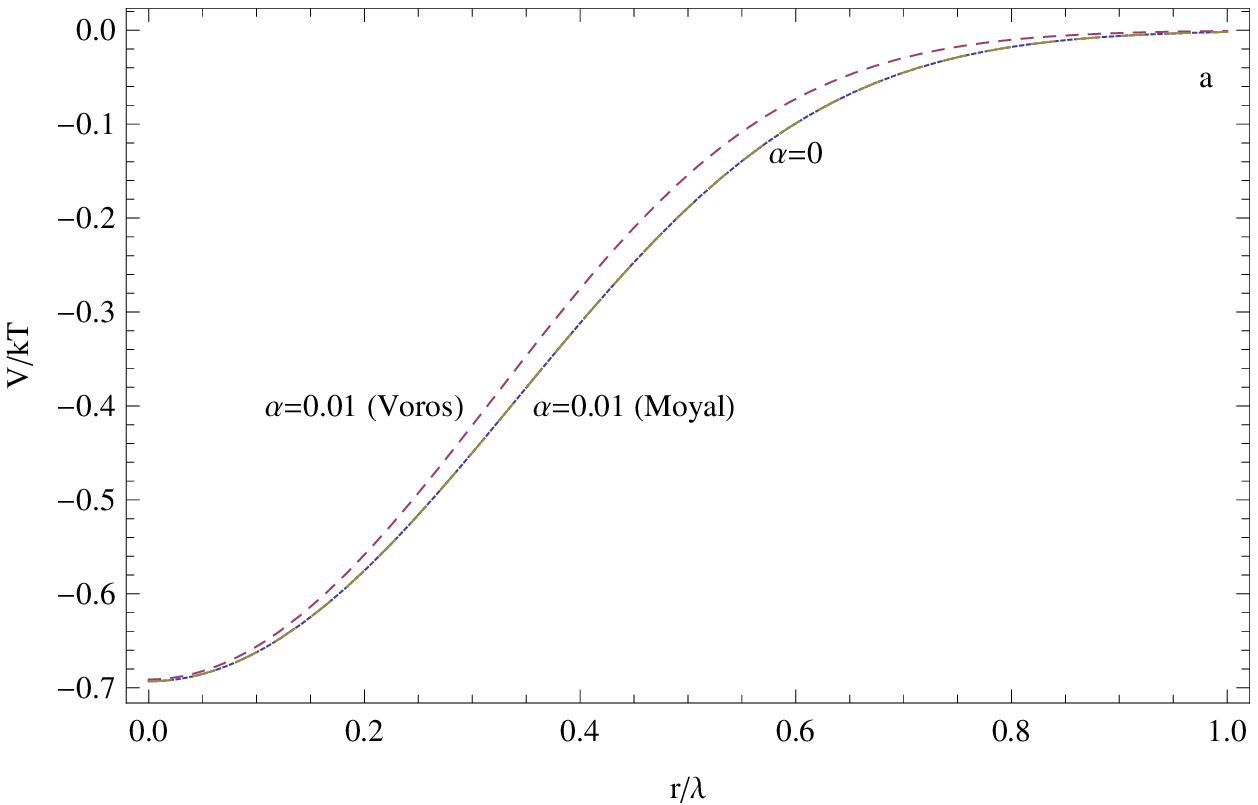, height=53mm}}
\put(80, 0){\epsfig{file=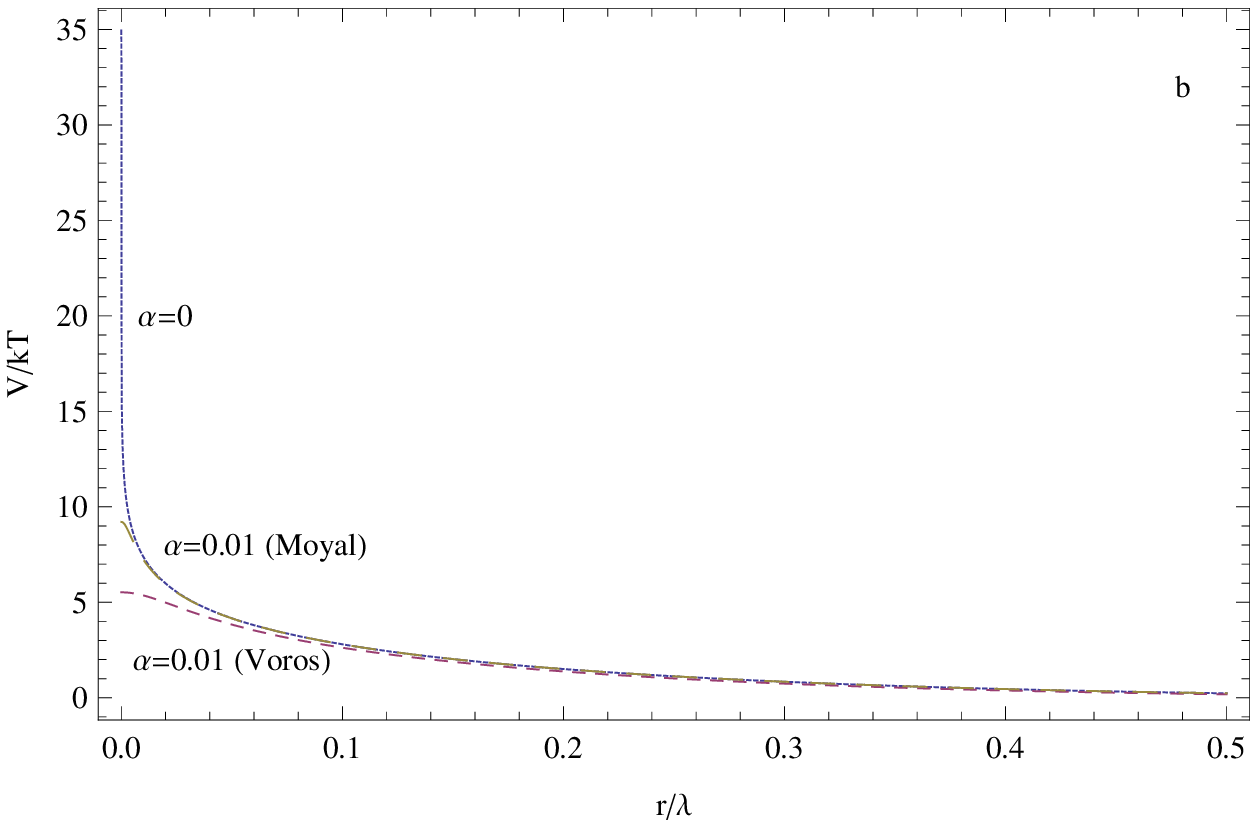, height=53mm}}
\end{picture}
\caption {The exchange potential in units of $k_{B}T$ for bosons (a) and fermions (b).  The solid curve represents the commutative result ($\alpha=0$), the short dashed curve is the result for the Voros twist and the long dashed curve the result for the Moyal twist.  The value of $\alpha=0.01$ is below the temperature $T_1$.}
\end{figure}

\setlength{\unitlength}{1mm}
\begin{figure}
\begin{picture}(53,53)
\put(0, 0){\epsfig{file=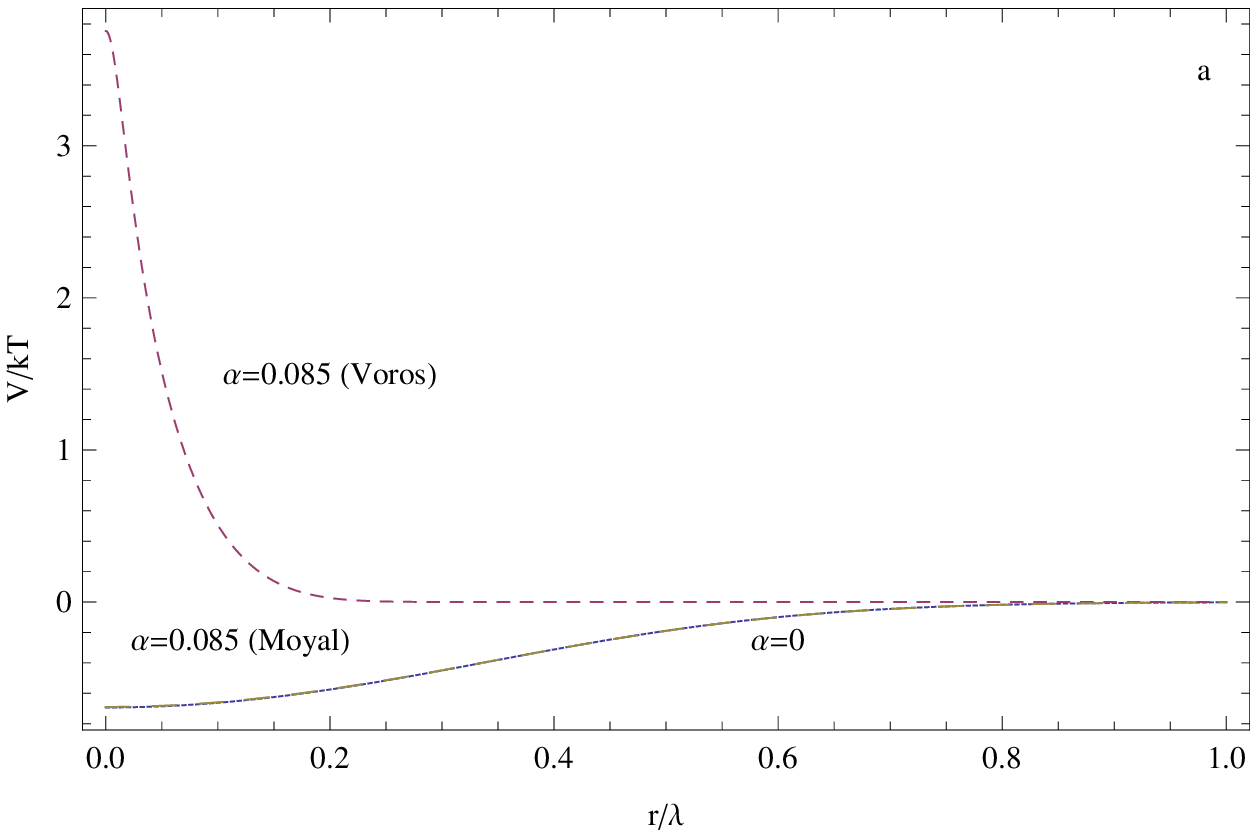, height=53mm}}
\put(80, 0){\epsfig{file=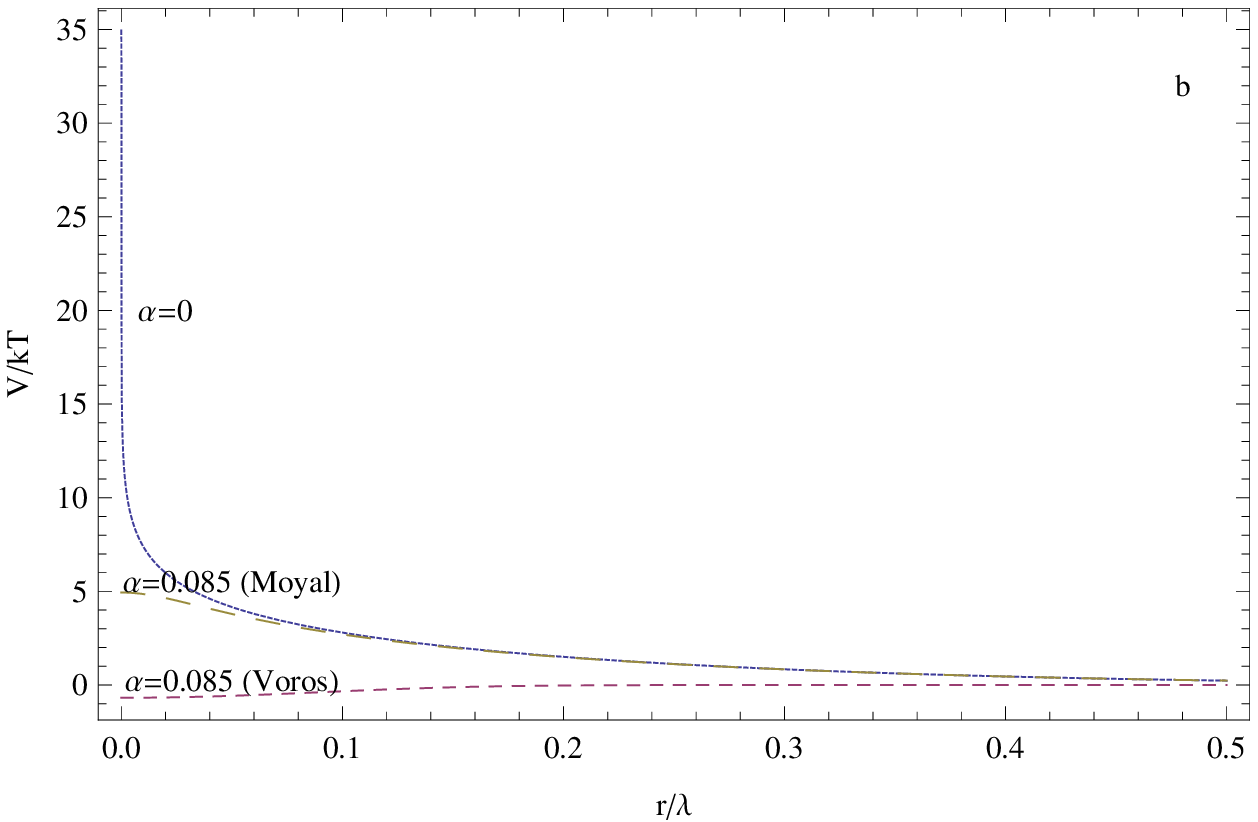, height=53mm}}
\end{picture}
\caption {The exchange potential in units of $k_{B}T$ for bosons (a) and fermions (b).  The solid curve represents the commutative result ($\alpha=0$), the short dashed curve is the result for the Voros twist and the long dashed curve the result for the Moyal twist.  The value of $\alpha=0.085$ is between the temperatures $T_1$ and $T_2$.  The Voros twist result exhibits for (a) fermionic rather than bosonic and (b) bosonic rather than fermionic behaviour.}
\end{figure}

\section*{Acknowledgement} The authors would like to thank the referees for very useful comments.




\end{document}